\documentclass[]{article}
\usepackage[utf8]{inputenc}
\usepackage{geometry}
\usepackage{float}
\usepackage{amsmath}
\usepackage{amsthm}
\usepackage{amssymb}
\usepackage{setspace}
\usepackage{enumitem}
\makeatletter

\theoremstyle{definition}
\newtheorem{defn}{\protect\definitionname}
\theoremstyle{plain}
\newtheorem{lyxalgorithm}{\protect\algorithmname}
\theoremstyle{remark}
\newtheorem{rem}{\protect\remarkname}
\theoremstyle{plain}
\newtheorem{thm}{\protect\theoremname}
\theoremstyle{definition}
\newtheorem{example}{\protect\examplename}
\theoremstyle{plain}

\theoremstyle{plain}

\providecommand{\algorithmname}{Algorithm}
\providecommand{\definitionname}{Definition}
\providecommand{\examplename}{Example}
\providecommand{\lemmaname}{Lemma}
\providecommand{\propositionname}{Proposition}
\providecommand{\remarkname}{Remark}
\providecommand{\theoremname}{Theorem}

\usepackage[style=apa,backend=biber,citestyle=authoryear,sorting=nty, natbib]{biblatex}

\usepackage{tikz}
\usetikzlibrary{arrows}

\makeatother

\begin{document}

\title{House-Swapping with Objective Indifferences}
\author{Will Sandholtz and Andrew Tai \\UC Berkeley}
\date{June 2023}

\maketitle

\begin{abstract}
We study the classic house-swapping problem of \citet{shapley_cores_1974} in a setting where agents may have ``objective'' indifferences, i.e., indifferences that are shared by all agents. In other words, if any one agent is indifferent between two houses, then all agents are indifferent between those two houses. The most direct interpretation is the presence of multiple copies of the same object. Our setting is a special case of the house-swapping problem with general indifferences. We derive a simple, easily interpretable algorithm that produces the unique strict core allocation of the house-swapping market, if it exists. Our algorithm runs in square-polynomial time, a substantial improvement over the cubed time methods for the more general problem. 
\end{abstract}


\section{Introduction}

The house-swapping problem originally studied by \citet{shapley_cores_1974} assumes that agents have a strict preference ordering over the set of the agents' houses. Implicitly, all houses are distinct. As \citet{roth_weak_1977} show, in this setting the strict core is always non-empty and consists of a single allocation, which can identified using the Top Trading Cycles algorithm (TTC).

In the more general setting where agents' preference rankings may contain indifferences, the strict core may be empty. Moreover, when the strict core is non-empty, it may contain multiple allocations. \citet{quint_houseswapping_2004} devised an algorithm, Top Trading Segmentation (TTS), that finds a strict core allocation, when it exists. \citet{alcalde-unzu_exchange_2011} devise Top Trading Absorbing Sets (TTAS) which finds the strict core when it exists and the weak core otherwise. They leave computational complexity of their algorithm as an open question. \citet{jaramillo-manjunath} also solve the general indifference problem with Top Cycle Rules (TCR), which has complexity $O(n^6)$. \citet{aziz-keijzer} present Generalized Absorbing Top Trading Cycle (GATTC), generalizing TTAS and TCR and show that TTAS has exponential time complexity. \citet{plaxton} develops a different mechanism to produce a strict core allocation with time complexity $O(n^3)$.

We study a more structured problem, where any indifferences are shared across all agents. We use the phrase ``objective indifferences'' to describe this setting. Conversely, we use the phrase ``subjective indifferences'' to describe indifferences that are not necessarily shared by all agents. Objective indifferences are the leading case of indifferences, since many objects we encounter in daily life are commodified. This additional structure enables us to develop a simple algorithm to find the strict core, when it exists, with time complexity $O(n^2)$.

Our setting can be thought of as an intermediate case between the original \citeauthor{shapley_cores_1974} setting and the general setting studied first by \citeauthor{quint_houseswapping_2004}. With objective indifferences, as in the house-swapping problem with subjective indifferences, the strict core may be empty. However, when the strict core is non-empty it contains a unique allocation. We propose a simple algorithm that finds the strict core allocation of a house-swapping market with objective indifferences in square-polynomial time. This algorithm is faster than the polynomial time algorithms that are needed for house-swapping markets with subjective indifferences.


\section{Model}

Let $I=\{1,2,...,I\}$ be a set of agents, each of whom is endowed with a house. Let $H=\{1,2,...,H\}$ be the set of possible house types in the market. Note that $H<I$ implies that some agents were endowed with houses of the same type. The endowment function $E:I\to H$ maps each agent to the house type he was endowed with.

Each agent $i\in I$ has strict preferences $\succsim_{i}$ over $H$. Implicitly, all agents are indifferent between two houses of the same type. We use $\succsim=\{\succsim_{1},\succsim_{2},...,\succsim_{I}\}$ to denote the preference profile of all agents.

An allocation $\mu$ is a function $\mu:I\to H$ such that $|\mu^{-1}(h)|=|E^{-1}(h)|$ for all $h\in H$. That is, $\mu(i)=h$ means agent $i$ is assigned a house of type $h$, and the number of agents who are allocated to a house type is equal to the supply of it.

The house-swapping market is summarized as the tuple $(I,H,E,\succsim)$. We are interested in whether the strict core exists.

\begin{defn}
An (sub-)allocation $\mu$ is \textbf{feasible} for a coalition of agents $I'\subseteq I$ if $\left|\mu^{-1}(h)\right|=\left|E^{-1}(h)\cap I'\right|$ for any $h \in E(I')$. That is, the quantity of each house type required in the (sub-)allocation is the same as the quantity in the coalition's endowment.
\end{defn}

\begin{defn}
A (feasible) allocation $\mu$ is in the \textbf{strict core} of the house-swapping market $(I,H,E,\succsim)$ if there is no coalition $I'\subseteq I$ and no sub-allocation $\mu'$ such that: 

\begin{enumerate}
\item $\mu'$ is feasible for $I'$
\item $\mu'(i)\succsim_{i}\mu(i)$ for all $i\in I'$ 
\item $\mu'(i)\succ_{i}\mu(i)$ for at least one $i\in I'$ 
\end{enumerate}
\end{defn}

We derive an algorithm that finds the strict core of a house-swapping market $(I,H,E,\succsim)$ when it exists.


\section{Directed Graphs}

Before proceeding to our main results and the algorithm, we review some useful concepts related to directed graphs. The definitions are standard, and a familiar reader may skim this section.

A \textbf{directed graph} is given by $D(V,E)$ where $V$ is the set of vertices and $E$ is the set of arcs. An \textbf{arc} is a sequence of two vertices $(v,v')$. We allow for arcs of the form $(v,v)$, which we call \textbf{self-loops}. A $(v_{1},v_{k})$\textbf{-path} is a sequence of vertices $(v_{1},v_{2},...,v_{k})$ where each $v_{i}$ is distinct and $(v_{i-1},v_{i})\in E$ for all $i=2,3,...,k$. A \textbf{cycle} is a path where $v_{1}=v_{k}$ is the only repeated vertex. A \textbf{sink} of a directed graph is a vertex $v$ such that $(v,v') \notin E$ for all $v' \in V$.

A \textbf{strongly connected component (SCC)} of a directed graph $D(V,E)$ is a maximal set of vertices $S\subseteq V$ such that for all distinct vertices $v,v'\in S$, there is both a $(v,v')$-path and a $(v',v)$-path. By convention, there is always a path from $v$ to itself, regardless of whether $(v,v)\in E$. The collection of strongly connected components of a directed graph forms a partition of $V$. (To see this, note that the definition of an SCC implies that a vertex can be in exactly one SCC.)

The \textbf{condensation} of a directed graph $D(V,E)$ is the directed graph $D(V^{SCC},E^{SCC})$ where $V^{SCC}$ is the set of SCCs of $D(V,E)$ and $(S,S') \in E^{SCC}$ if and only if there exist $v \in S$ and $v' \in S'$ such that $(v,v') \in E$. In other words, it is the arc-contraction of $D$ on each SCC -- replace each SCC with a single vertex, and keep any arcs between SCCs. Condensations of directed graphs are always acyclic. 

A \textbf{topological ordering} of a directed acyclic graph $D(V,E)$ is a total order $\leq$ of the elements of $V$ such that if $(v,v') \in E$, then $v \leq v'$. A directed graph has a topological ordering if and only if it is acyclic.\footnote{See \citet{korte-vygen_combinatorialopt_2008}, Section 2.2.} It is immediate that the vertex with the highest topological ordering is a sink.


\section{Results}

In this section, we give our algorithm to determine whether a strict core of a market $\left(I,H,E,\succsim\right)$ exists and to find it when it does. First, we define a function $B_{i}$ that denotes the $i$'s most preferred house type among a subset of house types. Let $B_{i}:I\times\mathcal{P}(H)\to H$ be given by $B_{i}(H')=h$ if $h\succsim_{i}h'$ for all $h'\in H'$.

We now give our algorithm.
\begin{lyxalgorithm}
\label{alg:StrictCore}House Top Trading Segments (HTTS)
\begin{description}[font=\normalfont]

\item[Step 1.] Let $R_{1}=H$. Construct the directed graph $D_{1}=D(R_{1},E_{1})$ where $(h,h')\in E_{1}$ if $B_{i}(R_{1})=h'$ for some $i\in E^{-1}(h)$. That is, draw an arc $(h,h')$ exists if an owner of $h$ top-ranks $h'$ among all house types $R_{1}=H$. Find an SCC $H_{1}$ of $D_{1}$ with no outgoing arcs; i.e., for any $h\in H_{1}$ and $h'\notin H_{1}$, $(h,h')\notin E_{1}$.\footnote{There always exists an SCC with no outgoing arcs. To see this, consider the condensation (contract each SCC to a single vertex). The result is a directed acyclic graph, which has at least one sink. The sink is the (contracted) desired SCC with no outgoing arcs. Note that there may be multiple SCCs with no outgoing arcs. If so, pick any arbitrarily.} We call $H_{1}$ a ``house top trading segment''.

\begin{enumerate}
\item Let $I_{1}=E^{-1}(H_{1})$. For all $i\in I_{1}$, set $\mu(i)=B_{i}(R_{1})$. That is, assign every agent endowed with a house in $H_{1}$ to his favorite house (also in $H_{1}$).
\item Check that $\mu$ is feasible for $I_{1}$. If so, proceed to part c. Otherwise, stop.
\item Let $R_{2}=R_{1}\setminus H_{1}$. If $R_{2}=\emptyset$, stop; otherwise, proceed to Step 2.
\end{enumerate}

\item[Step $d$.] Construct the directed graph $D_{d}=D(R_{d},E_{d})$ where $(h,h')\in E_{d}$ if $B_{i}(R_{d})=h'$ for some $i\in E^{-1}(h)$. Find an SCC $H_{d}$ of $D_{d}$ with no outgoing arcs. 

\begin{enumerate}
\item Let $I_{d}=E^{-1}(H_{d})$. For all $i\in I_{d}$, set $\mu(i)=B_{i}(R_{d})$. That is, assign each agent in $I_{d}$ to his favorite remaining house. Since $H_{d}$ has no outgoing arcs, this house is also in $H_{d}$.
\item Check that $\mu$ is feasible for $I_{d}$. If so, proceed to part $c$. Otherwise, stop.
\item Let $R_{d+1}=R_{d}\setminus H_{d}$. If $R_{d+1}=\emptyset$, stop; otherwise, proceed to Step $d+1$.
\end{enumerate}
\end{description}
\end{lyxalgorithm}

\begin{rem}
Note that at each step, house types are removed, and thus agents owning them are also removed. Since there are finitely many house types $H$, the algorithm terminates in finite time.
\end{rem}
\begin{rem}
At part $b$ of each step, $\mu$ is feasible for $I_{d}$ if and only if for each $h\in H_{d}$, $\left|E^{-1}(h)\cap I_{d}\right|=\left| \left\{ i:B_{i} H_{d})=h,i\in I_{d} \right\} \right|$. That is, the number of copies of $h$ available in $I_{d}$ is equal to the number of agents who top-rank $h$ among the remaining houses. Informally, ``supply equals demand.''
\end{rem}

The house top trading segments we find in each step are analogous to TTC trading cycles. At each step, agents ``point'' from their owned house to their favorite house. We then find the trading segment and execute the trades, if possible (``feasible''). For readers familiar with Quint and Wako (2004), these are modified versions of top trading segments.

\begin{thm}
\label{thm:MainThrm}Let $\left(I,H,E,\succsim\right)$ be a market.
\begin{enumerate}
\item The strict core exists if and only if Algorithm \ref{alg:StrictCore} terminates in part $c$ of a step. That is, each step's HTTS gives a feasible allocation, and the algorithm did not terminate in part $b$ of a step.
\item Algorithm \ref{alg:StrictCore} finds a strict core allocation, when one exists.
\item The strict core allocation is unique, when it exists.\footnote{Recall the definition of an allocation is a matching between agents and house types. The individual identities of the houses do not matter.}
\item Algorithm \ref{alg:StrictCore} has time complexity $O(|H|^2+|H||I|)$.
\end{enumerate}
\end{thm}

Before the proof of Theorem \ref{thm:MainThrm}, we give the following example to illustrate it and Algorithm \ref{alg:StrictCore}.

\begin{example}
\label{exa:mainEx}Consider the house-swapping market $\mathcal{}(I,H,E,\succsim)$
where 
\begin{align*}
 & I=\{1,2,3,4,5\}\\
 & H=\{h_{1},h_{2},h_{3},h_{4}\}\\
 & E(1)=h_{1},\;E(2)=E(3)=h_{2},\;E(4)=h_{3},\;E(5)=h_{4}
\end{align*}
and $\succsim=\{\succsim_{1},\succsim_{2},\succsim_{3},\succsim_{4},\succsim_{5}\}$
is given by 
\begin{align*}
h_{2} & \succ_{1}...\\
h_{1} & \succ_{2}...\\
h_{3} & \succ_{3}h_{2}\succ_{3}...\\
h_{4} & \succ_{4}...\\
h_{3} & \succ_{5}...
\end{align*}

\begin{enumerate}
\item \textit{Step 1}: Set $R_{1}=H$. Construct the directed graph $D(R_{1},E_{1})$ where $(h,h')\in E_{1}$ if $B_{i}(R_{1})=h'$ for some $i\in E^{-1}(h)$. That is, some owner of $h$ top ranks $h'$. There are two SCCs in $D(R_{1},E_{1})$: $\{h_{1},h_{2}\}$ and $\{h_{3},h_{4}\}$. Only $S=\{h_{3},h_{4}\}$ has no outgoing arcs. Then set $H_{1}=\{h_{3},h_{4}\}$ and $I_{1}=\{4,5\}$.
\begin{enumerate}
\item Assign $\mu(4)=h_{4};\mu(5)=h_{3}$.
\item Check that this is feasible for $I_{1}$. We have 
\begin{align*}
\left|E^{-1}(h_{3})\cap I_{1}\right|=\left|\{4\}\right| & =1\\
\left|\{i:B_{i}(H_{1})=h_{3},i\in I_{1}\}\right|=\left|\{5\}\right| & =1
\end{align*}
and likewise for $h_{4}$, so this is feasible.
\item Set $R_{2}=R_{1}\setminus H_{1}=\{h_{1},h_{2}\}$ and continue to Step 2.
\end{enumerate}

\item \textit{Step 2}: Construct the directed graph $D(R_{2},E_{2})$ where $(h,h')\in E_{2}$ if $B_{i}(R_{2})=h'$ for some $i\in E^{-1}(h)$. That is, some owner of $h$ top ranks $h'$ among the remaining houses $R_{2}=\{h_{1},h_{2}\}$. The entire graph forms an SCC, so set $H_{2}=\{h_{1},h_{2}\}$ and $I_2 = \{1,2,3\}$.
\begin{enumerate}
\item Assign $\mu(1)=h_{2};\mu(2)=h_{1};\mu(3)=h_{2}$.
\item Check that this is feasible for $I_{2}$ (it is).
\item Set $R_{3}=R_{2}\backslash H_{2}=\emptyset$. So the algorithm terminates.
\end{enumerate}
\end{enumerate}

Therefore, a House Top Trading Segmentation of $H$ is given by 
\[
\mathcal{H}=\big\{ H_{1}=\{h_{3},h_{4}\},H_{2}=\{h_{1},h_{2}\}\big\}.
\]

\begin{figure}[H]
\centering %
\begin{minipage}[t]{0.49\textwidth}%
\centering \begin{tikzpicture}
  \centering
  \tikzset{vertex/.style = {shape=circle,draw,minimum size=1em}}
  \tikzset{edge/.style = {->,> = latex'}}
  \node[vertex] (a) at (0,2) {$h_1$};
  \node[vertex] (b) at (2,2) {$h_2$};
  \node[vertex] (c) at (0,0) {$h_3$};
  \node[vertex] (d) at (2,0) {$h_4$};
  
  \draw[edge] (a) to[bend left] (b);
  \draw[edge] (b) to[bend left] (a);
  \draw[edge] (b) to (c);
  \draw[edge] (c) to[bend left] (d);
  \draw[edge] (d) to[bend left] (c);

  \draw[densely dashed, gray] (-.75,1.25) rectangle (2.75,2.75);
  \draw[densely dashed, red] (-.75,-.75) rectangle (2.75,.75);
  \node[] at (3.25,0) {$H_1$};
  \node[] at (3.25,2) {$R_2$};
  \node[] at (1,-1.5) {Step 1};
  \end{tikzpicture}%
\end{minipage}\hfill{}%
\begin{minipage}[t]{0.49\textwidth}%
\centering \begin{tikzpicture}
  \centering
  \tikzset{vertex/.style = {shape=circle,draw,minimum size=1em}}
  \tikzset{edge/.style = {->,> = latex'}}
  \node[vertex] (a) at (0,0) {$h_1$};
  \node[vertex] (b) at (2,0) {$h_2$};
  
  \draw[edge] (a) to[bend left] (b);
  \draw[edge] (b) to[bend left] (a);
  \draw[edge] (b) to[loop above] (b);

  \draw[densely dashed, red] (-.75,-.75) rectangle (2.75,.75);
  \node[] at (3.25,0) {$H_2$};
  \node[] at (1,-1.5) {Step 2};
  \end{tikzpicture}%
\end{minipage}\caption{Applying Algorithm \ref{alg:StrictCore} to Example \ref{exa:mainEx}.}
\end{figure}
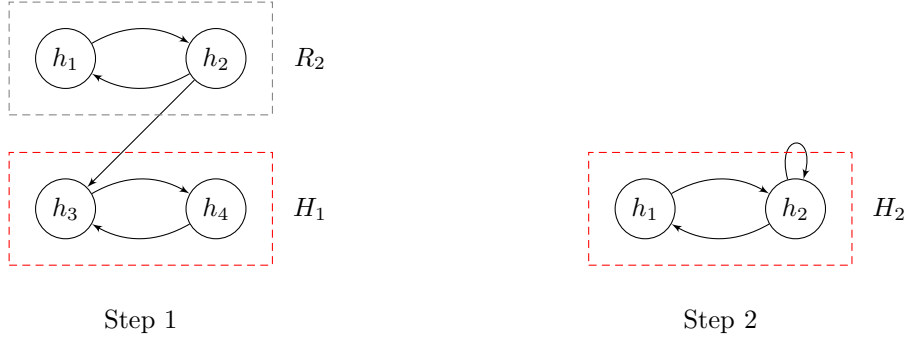

By Theorem \ref{thm:MainThrm}, the unique strict core of this
market is given by
\begin{align*}
\mu(1) & =h_{2}\\
\mu(2) & =h_{1}\\
\mu(3) & =h_{2}\\
\mu(4) & =h_{4}\\
\mu(5) & =h_{3}
\end{align*}

\end{example}


\subsection{Proof of Theorem \ref{thm:MainThrm}}

The proofs for the strict core claims unsurprisingly follow Gale's proof for TTC. The first key insight is that by focusing on house types as nodes (instead of agents), we ensure that we remove all copies of a house at the same time. This lets us easily deal with objective indifferences. The second key insight is that when we assign houses within an SCC without outgoing arcs, we assign a set of houses and their owners at the same time.

\begin{proof}[Proof of Claim 2.]
Let $\left(I,H,E,\succsim\right)$ be a market, and let $\mu^{HTTS}$ be the allocation produced by Algorithm \ref{alg:StrictCore}. That is, the algorithm terminated in part $c$ of some step.

We first argue that $\mu^{HTTS}$ is indeed a feasible allocation. At each step, we arrive at a house trading segment $H_{d}$. Note that $H_{d}$ has no outgoing arcs in $D_{d}$. Thus all agents endowed with a house $h\in H_{d}$ (denoted $I_{d}$) top-rank a house in $H_{d}$ from among the remaining houses. By our assumption that Algorithm 1 terminated in part c (and not part b) of some step, we know that $\mu^{HTTS}$ is feasible for $I_{d}$. Part $c$ of this step removes $H_{d}$ and thus $I_{d}$ from further consideration. Thus $\{H_{1},...,H_{d},...,H_{K}\}$ and $\{I_{1},...,I_{d},...,I_{K}\}$ partition the house types and agents, respectively. If $\mu$ is feasible for each $I_{d}$, then it is feasible for $I$.

Toward a contradiction, suppose there is a blocking coalition $I'$ and sub-allocation $\mu'$.

For at least one agent $i\in I$, $\mu'(i)\succ\mu^{HTTS}(i)$. Consider the step $d$ at which $i$ was assigned in Algorithm \ref{alg:StrictCore}. By construction, $\mu^{HTTS}(i)=B_{i}(H_{d})=B_{i}\left(\cup_{d'\geq d}H_{d'}\right)$. So it must be that $\mu'(i)\in\cup_{d'<d}H_{d'}$. Feasibility of $\mu'$ implies that there is some $i'\in I_k$ for $k < d$ such that $\mu'(i')\in\cup_{k' > k}H_{k'}$. But then $\mu'(i')\prec\mu^{HTTS}(i')$, so this is not a blocking coalition. In other words, for $\mu'(i)\succ\mu^{HTTS}(i)$, $i$ must be assigned to a house from an earlier segment. But then an agent from an earlier segment must be assigned to a house from a later segment, which is strictly dispreferred.
\end{proof}

\begin{proof}[Proof of Claim 3.]
Let $\left(I,H,E,\succsim\right)$ be a market, and let $\mu^{HTTS}$ be the allocation produced by Algorithm \ref{alg:StrictCore}. That is, the algorithm terminated in a part $c$. Let $\mu'$ be another strict core allocation. We again show $\mu'=\mu^{HTTS}$ by strong induction on the number of steps in HTTS.
\begin{description}[font=\normalfont]
\item[Base claim.]
Consider $H_{1}$ and $I_{1}$. We have $\mu^{HTTS}(i)\succsim_{i}\mu'(i)$ for all $i\in I_{1}$, since every $i\in I_{1}$ receives his favorite house. Since $\mu^{HTTS}$ is feasible for $I_{1}$ and $\mu'$ is in the strict core, we must also have $\mu'(i)\succsim_{i}\mu^{HTTS}(i)$ for all $i\in I_{1}$. (Otherwise $I_{1}$ can form a blocking coalition with sub-allocation $\mu^{HTTS}|_{I_{1}}$.) But then $\mu^{HTTS}(i)=\mu'(i)$ for $i\in I_{1}$.
\item[Claim $d$.] Assume $\mu^{HTTS}(i)=\mu'(i)$ for all $i\in I_{1}\cup\cdots\cup I_{d-1}$. Then $\mu'(I_{d})\subseteq \cup_{d'\geq d}H_{d'}$. That is, the houses assigned to agents in $I_{d}$ are drawn from the houses that remain after step $d-1$. By construction, we have $\mu^{HTTS}(i)\succsim_{i}h$ for any $h\in\cup_{d'\geq d}H_{d'}$ for all $i\in I_{d}$, so we have $\mu^{HTTS}(i)\succsim_{i}\mu'(i)$ for $i\in I_{d}$. Since $\mu^{HTTS}$ is feasible for $I_{k}$ and $\mu'$ is in the strict core, we must have $\mu'(i)\succsim_{i}\mu^{HTTS}(i)$ for all $i\in I_{d}$. But then $\mu^{HTTS}(i)=\mu'(i)$ for $i\in I_{d}$.
\end{description}
\end{proof}

\begin{proof}[Proof of Claim 1.]
We now have that $\mu^{HTTS}$ is the unique strict core allocation, when it exists. Thus, if $\mu^{HTTS}$ is not feasible, there is no strict core allocation.
\end{proof}

\begin{proof}[Proof of Claim 4.]

We apply Tarjan's algorithm \citep{tarjan_depth-first_1972}. For any directed graph $G = D(V,E)$, the order in which Tarjan's algorithm returns the SCCs of $G$ is a reverse topological ordering of the condensation $G^{SCC} = D(V^{SCC},E^{SCC})$ of $G$.\footnote{See \citet{korte-vygen_combinatorialopt_2008}, Section 2.3}. Concretely, suppose $\mathcal{S} = \{S_1,S_2,...,S_\ell\}$ is the set of SCCs of $G$ in the order in which they were returned by Tarjan's algorithm (i.e., $S_1$ is the first SCC returned, $S_2$ is the second, etc.). Then $S_1$ must be a sink of $G^{SCC}$. Therefore, $S_1$ is an SCC of $G$ with no outgoing arcs. 

At each step $d$ of Algorithm \ref{alg:StrictCore}, we perform two computations. First, we use Tarjan's algorithm to identify an SCC $H_d$ with no outgoing arcs.\footnote{We need not find all SCCs. The first SCC returned by Tarjan's algorithm will suffice.} Tarjan's algorithm has time complexity $O(|H| + |I|)$. Second, we check whether the strict core allocation is feasible for $I_d = E^{-1}(H_d)$. That is, for each $h \in H_{d}$, we check $\left|E^{-1}(h)\cap I_{d}\right|=\left| \left\{ i:B_{i}(H_{d})=h,i\in I_{d} \right\} \right|$. This has time complexity $O(|H|)$. Therefore, each step of Algorithm \ref{alg:StrictCore} has time complexity $O(|H| + |I|)$.

Since Algorithm \ref{alg:StrictCore} terminates in at most $|H|$ steps, it has time complexity $O(|H|^2 + |H||I|)$.

\end{proof}


\section{Conclusion}

In this paper, we study the house-swapping problem in a setting where agents' preferences may contain ``objective indifferences.'' We assume that agents have strict preferences over a set of house types and that multiple agents may be endowed with copies of the same house type. We derive a square-polynomial time algorithm that finds the unique strict core allocation of a house-swapping market, if it exists. This is faster than the methods that are needed to find strict core allocations in the setting where agents are allowed to have subjective indifferences. Moreover, our algorithm is interpretable as a series of ``house top trading segments'', which are analogous to top trading cycles. The condition for the non-emptiness of the strict core is readily interpretable -- within each house top trading segment, supply and demand for each house type are equal.


\nocite{*}
\printbibliography

\end{document}